\documentclass[12pt]{amsart}
\input{epsf}             

\newcommand{\dd}{{\rm d}} 

\newcommand{\R}{{\mathbb R}}

\begin{document}

\title{Refractive gravitational waves and quantum fluctuations}  
\author{John W.\ Barrett }
\address{School of Mathematical Sciences, University
Park, Nottingham NG7 2RD, UK}

\begin{abstract} Refractive gravitational waves are a generalisation of impulsive waves on a null hypersurface in which the metric is
discontinuous but a weaker continuity condition for areas holds. A simple example of a plane wave is examined in detail and two arguments are given that this should be considered a solution of Einstein's vacuum field equations. The study of these waves is motivated by quantum gravity, where the refractive plane waves are considered as elementary quantum fluctuations and the `area geometry' of a null hypersurface plays a primary role. 
\end{abstract}

\maketitle 

\section{Introduction} In Euclidean geometry, it is hard to see how the measurement of the areas of two-dimensional surfaces could be taken to be the fundamental basis of the theory. In four-dimensional Lorentzian geometry however, areas of surfaces play an important role in various physical phenomena, the most famous example being the area of a black hole event horizon and its role in quantum mechanical phenomena associated with the black hole. It is noteworthy in this example that the event horizon is a null hypersurface, and this allows the area of spatial surfaces in the horizon to have a more fundamental significance. Of course, null hypersurfaces do not occur in Euclidean geometry.

In this paper I would like to introduce a very simple space-time in which the metric is discontinuous at a null hypersurface, but `areas' are continuous and play a fundamental role in the geometry. This is motivated by the study of models of quantum gravity, but it appears to be interesting in classical general relativity as an example of a new class of idealised gravitational wave spacetimes.

\section{Refractive waves}

An impulsive gravitational wave $W$ has the structure, locally, of two vacuum spacetimes $(M,g)$, $(M',g')$, for which the manifolds $M$ and $M'$ intersect along a common boundary $N$, so that $W=M\cup M'$ and $M\cap M'=N$. The boundary $N$ is a null hypersurface for both $(M,g)$ and $(M',g')$, and the two induced metrics on $N$ determined by the metrics $g$ and $g'$ are equal. The Einstein equations for $W$ are the condition that a certain part of the affine connections for $M$ and $M'$ agree along $N$ \cite{P,MS}.

However a null hypersurface in a spacetime has a weaker structure than its $(--0)$ signature metric. Every two-dimensional surface in $N$ has an area determined by the metric. Therefore the junction condition for the impulsive gravitational wave that the two induced metrics agree can be replaced by the weaker condition that the area of any surface in $N$ measured with $g$ agrees with the area measured with $g'$.
This gives a more general notion of localized gravitational wave which I shall call a refractive gravitational wave.\footnote{
The two spacetimes $M$ and $M'$ are also required to be causally complementary. This means that if, locally, the future of a point $p\in N$ is contained in $M$, then the past is in $M'$, or vice-versa.}
 
The area matching condition can be rephrased in a differential manner as follows. Assuming the null hypersurface $N$ is oriented, and the spacetime is time-oriented, then there is a two-form $\Omega$ on $N$
which is defined to give the area of each spacelike 2-surface in the hypersurface by integration. This area has a sign depending on the orientation of the surface. More explicitly, the tangent space $T$ at a point of $N$ has a one-dimensional subspace $K$ orthogonal to $T$. The quotient $T/K$ has a non-degenerate metric and an orientation and thus a volume form (giving the area). This volume form pulled back to $T$ defines $\Omega$.  The area matching condition is then equivalent to the condition that if $\Omega$ is the 2-form on $N$ determined by $g$ and $\Omega'$ the 2-form on $N$ determined by $g'$, then $\Omega=\Omega'$.

The form $\Omega$ carries the information about the null directions: a vector $k$ is null if and only if
$$ \Omega(k,n) =0 \qquad \text{for all $n$ tangent to $N$.}$$ 
Therefore the area matching condition implies that the null directions in $N$ determined by $g$ and $g'$ agree. At each point the space of metrics of signature $(--0)$ on the tangent space is five dimensional, but the space of 2-forms is three dimensional. This means there are two parameters at each point for the mismatch of $g$ and $g'$. 

To summarise, in a refractive wave spacetime the two induced metrics determined by $g$ and $g'$ will not agree, so that the lengths of curves in $N$ are not uniquely determined, but both areas of surfaces and null directions do agree, defining an \textit{area geometry} for the hypersurface $N$.

Discontinuous metrics are familiar from optics. The speed of light in an optical medium can be coded into a Riemannian metric in three-dimensional space which gives the time taken for light to travel along a path. At a boundary $N$ between two media, the induced metrics $g$ and $g'$ do not agree (of course, $N$ is not null in this example). A ray of light is refracted by the familiar law of sines of optics. If the unit tangent vector of the ray incident to $N$ is $\xi$, and the unit tangent vector on the other side of $N$ is $\xi'$, both pointing along the ray in the same direction, the law of refraction can be written in the form
\begin{equation} g(\xi,n)=g'(\xi',n) \qquad \text{for all $n$ tangent to $N$.}\label{refraction}  \end{equation}

In the refractive wave spacetime $W$, the variational principle for the geodesic equation can be formulated in entirely the same way for any geodesic not tangent to the null hypersurface $N$. A timelike, or spacelike, geodesic is a stationary curve for the length functional.
The same law of refraction (\ref{refraction}) expresses the stationary condition for the total length of the curve. One can easily see that there is always a unique continuation of a geodesic which is incident on $N$ (the phenomenon of total internal reflection is not necessary for null hypersurfaces in Minkowski signature, as a solution to (\ref{refraction}) always exists).

The main problem with the refractive waves is to determine what the Einstein equations could be. In general this is an open problem but a solution is suggested in the next section for the simplest case of a plane wave.

\subsection{Refractive plane waves}

For these examples, the manifold is $\R^4$, with $N$ a hyperplane $u=0$ cutting it into two manifolds with boundary, $M$ and $M'$. These two pieces have metric tensors $g$ and $g'$ respectively which are constant, meaning that the metric coefficients do not depend on the coordinates of $\R^4$. The metric on the whole of $\R^4$ can thus be written as
$$g_{ab}\Theta(u) + g'_{ab}(1-\Theta(u)).$$
The resulting spacetime $W_p(g,g')$ is a refractive gravitational wave if $g$ and $g'$ are causally complementary and $\Omega=\Omega'$. The spacetime can be constructed from Minkowski space by cutting it in two along a null hyperplane through the origin, and reassembling the pieces by an orientation-preserving and time-orientation preserving linear isomorphism $N\to N$ which preserves $\Omega$.
 
Since the metric tensor is not continuous on $W_p$, the Einstein field equations can not be applied to it in the usual formalism. Nevertheless it has the characteristics of a solution of these equations, and I propose that it be considered as a singular wave solution of the Einstein equations. I will present two pieces of evidence for this. The first is considering the linearised equations, and the second is considering an example of a continuous sandwich wave which approximates a refractive wave.

\subsection{Linearised appoximation} Let $l$ be a first-order perturbation on the Minkowski metric $\eta$. The linearised Einstein equation
$$
\eta^{ae}(\partial_a\partial_b l_{de}+\partial_e\partial_d l_{ba}
-\partial_a\partial_e l_{db} - \partial_d\partial_b l_{ae})
-\eta^{ae}\eta^{fh}(\partial_a\partial_f l_{he}-\partial_a\partial_e l_{hf})\eta_{bd}=0$$
makes sense for discontinuous $l$. In the example here, the linearised limit is $g=\eta+\epsilon h$, $g'=\eta+\epsilon h'$ in the limit of small $\epsilon\in\R$, so that
$$l_{ab}(u)=h_{ab}\Theta(u) + h'_{ab}(1-\Theta(u))$$
where $u$ is a null coordinate in Minkowski space, $N$ is the hypersurface $u=0$, and $\Theta$ is the Heaviside function. The constant tensors $h$ and $h'$ define constant perturbations of the Minkowski metric on each side of $N$.

Let $k_a=\partial_a u$ be the normal to $N$, a null vector. The linearised Einstein equations for this example are equivalent to the condition that $H_{ab}=h_{ab}-h'_{ab}$ satisfies
$$ H_{ab}k^a=\frac 1 2 H_{ac}\eta^{ac} k_b.$$
This is equivalent to the condition that the trace of $H_{ab}$ restricted to any spacelike 2-surface in $N$ is zero. Since the trace is a perturbation of a determinant, this means that the linearised perturbations of areas for $l$ and $l'$ are equal.  The conclusion is that the matching condition on $\Omega$ is equivalent to the linearised Einstein equations.

In this linearised example, the Weyl tensor is proportional to $\delta'(u)$ and there is consequently no geodesic deviation. This is consistent with the refraction equations on (the non-linear) $W_p$, which do not show geodesic deviation. 

\subsection{Sandwich wave appoximation} The second consideration is a sandwich plane wave metric\cite{RI}. This is a continuous metric which is a solution to the Einstein equations and has non-zero curvature for $-a<u<a$.
The metric is
$$\dd u\, \dd v -p(u)^2\, \dd x^2 - q(u)^2\, \dd y^2.$$
This has one Einstein equation,
$$\frac{p''}p +\frac{q''}q=0.$$
This can be solved by taking an odd function $\alpha(u)$ which is zero for $u>a$ and $u<-a$, and solving $p''/p=\alpha$. As long as the solution satisfies $p>0$ for $-a<u<a$, setting $q(u)=p(-u)$ solves the Einstein equation.
Outside the interval $-a<u<a$, both $p$ and $q$ are linear functions of $a$ and it is well-known that this gives Minkowski space, with coordinate singularities where either $p$ or $q$ is equal to zero.

One can think of this solution as essentially two sandwich 
wave approximations to impulsive waves back-to-back, with cancelling strengths. However the gravitational effect is non-zero due to the displacement between their centres. In effect, this is a non-linear version of the $\delta'$ linearised solution noted earlier.

The geodesic equations show that for a geodesic passing through the wave (i.e., $u$ is not constant) $u$ is an affine parameter. The equations for $x$ and $y$ integrate to give 
$$p(u)^2\; \frac{\dd x}{\dd u}=\text{constant}, \qquad q(u)^2\; \frac{\dd y}{\dd u}=\text{constant}.$$ 
These equations lead to the remarkable fact that the velocity vector is propagated along the geodesic by the law of refraction alone. Let $N$ be the null hypersurface $u=a$ and $N'$ the hypersurface $u=-a$. Let $\xi$ and $\xi'$ be the velocity vectors for a geodesic at the values $u=a$ and $-a$. Then a calculation shows that the relation between $\xi$ and $\xi'$ is
$$ g(\xi,n)=g'(\xi',\phi(n)) \qquad \text{for all $n$ tangent to $N$},$$ 
where $\phi$ is the map which takes a point in $N$ to the point in $N'$ with the same $v$, $x$, $y$ coordinates. This is the same as the law of refraction (\ref{refraction}) if $\phi$ is used to identify $N$ with $N'$. 

The other important feature of the map $\phi$ is that it is area-preserving. Indeed, $p(u)q(u)\,\dd x\,\dd y=p(-u)q(-u)\,\dd x\,\dd y$, and the null directions are preserved by $\phi$.

Therefore the properties of this sandwich wave are very close to the properties of the refractive plane wave. The correspondence is to think of the two sides of the refractive plane wave as the two flat outside portions of the sandwich wave, $u>a$ and $u<-a$. Identifying these portions with $\phi$ gives exactly the refractive plane wave spacetime.
The only difference is that geodesics travelling through the sandwich wave suffer a finite displacement in the $x$, $y$ and $v$ coordinates. However even this difference can be rectified by considering a limit of sandwich waves in which $a\to0$. This could be done in many ways\footnote{The limiting metric is not a regular metric in the sense of Geroch and Traschen\cite{GT}, and so the limit of the Weyl curvature may depend on the limiting sequence. This limit of the curvature is not considered in this paper.} but a convenient choice is to take a sandwich wave given by $p=P(u)$ and make a family of solutions depending on $\lambda$ by $p(u,\lambda)=P(u\lambda)$. As $\lambda\to\infty$ the thickness of the wave goes to zero and the displacements of $(x,y,v)$ also go to zero. This means that, in the limit, any geodesic incident at $z\in N$ emerges at $\phi(z)\in N'$. This is the justification for using $\phi$ to identify the two exterior portions to give a refractive wave spacetime. The propagation of geodesics in this limit is then exactly as for the refractive wave.

\section{Quantum fluctuations}

The idea is to think of the refractive plane wave $W_p(g,g')$ as an elementary quantum fluctuation of the gravitational field. For a given area geometry (2-form $\Omega$) on a hypersurface $N$, there is a configuration space of refractive plane waves which can be thought of as the corresponding quantum fluctuations. Interpreting metric discontinuities as elementary fluctuations was suggested previously in \cite{RA}.

This arises in a concrete way in state sum models of quantum gravity. The quantum tetrahedron is a quantum-mechanical model of a Euclidean tetrahedron in which the areas of the four faces (triangles) are classical variables but the remaining parameters for a tetrahedron embedded in Euclidean space have quantum-mechanical fluctuations \cite{BC, BZSF, BAR}. In particular, there are six parameters for a Euclidean metric, which means that with four areas fixed there are a further two parameters for the shape of the tetrahedron which are fluctuating.

In \cite{BB}, these fluctuations were analysed for a tetrahedron in $\R^4$ with the puzzling conclusion that the remaining two parameters for the metric of the tetrahedron have an irreducible quantum uncertainty, i.e. they cannot also be given sharp values. Although this is natural from the quantum theory, the interpretation in terms of a space-time, in the case of Euclidean signature, remains obscure.

In the Lorentzian version of the theory\cite{BC2}, the possibility of null tetrahedra arises. In this case the fluctuations in the metric of the tetrahedron have an interpretation in terms of the refractive plane waves.

The metric for a null tetrahedron now has five parameters instead of six because its determinant vanishes. A null tetrahedron $T$ still has four areas for its faces, but these are now signed, with the sign depending on whether the triangle is past or future facing in the boundary of $T$, and satisfy one relation,
$$A_1+A_2+A_3+A_4=0.$$
Thus there are three independent parameters.
This data is the same as specifying a 2-form
$$\Omega=\alpha \dd v \wedge \dd x + \beta\dd x \wedge \dd y + \gamma\dd y \wedge \dd v$$
with constant coefficients $\alpha$, $\beta$, $\gamma$, which determines the areas by integration. Thus the areas of the faces in a null tetrahedon exactly parameterise the area geometries of $N$. 

In \cite{BC} the $|A_i|$ specify certain unitary irreducible representations of the Lorentz group. One can interpret the canonical quantum state in the tensor product of these four representations as  the quantum state for a flat metric $g$ which agrees with the area geometry $\Omega$. This gives a natural interpretation of the metrics which arise in the asymptotic limit\cite{BW}: for a null tetrahedron there is a refractive plane wave between the two 4-simplexes on either side of the tetrahedron.

\end{document}